\documentclass[letterpaper,twocolumn,10pt]{article}
\usepackage{usenix2019_v3}

\usepackage{amsmath}
\usepackage{tikz}
\usepackage{placeins}

\begin{document}

\date{}

\title{\Large \bf Extraction of Secrets from 40nm CMOS Gate Dielectric Breakdown Antifuses\\
by FIB Passive Voltage Contrast}

\author{
{\rm Andrew D. Zonenberg}\\
IOActive
\and
{\rm Antony Moor}\\
IOActive
\and
{\rm Daniel Slone}\\
IOActive
\and{\rm Lain Agan}\\
IOActive
\and
{\rm Mario Cop}\\
IOActive
} 

\maketitle

\begin{abstract}
CMOS one-time-programmable (OTP) memories based on antifuses are widely used for storing small amounts of data (such as
serial numbers, keys, and factory trimming) in integrated circuits due to their low cost, requiring no
additional mask steps to fabricate.

Device manufacturers and IP vendors have claimed for years that antifuses are a ``high security" memory which is
significantly more difficult for an attacker to extract data from than other types of memory, such as Flash or mask ROM
- however, as our results show, this is untrue.

In this paper, we demonstrate that data bits stored in a widely used antifuse block can be extracted by a semiconductor
failure analysis technique known as passive voltage contrast (PVC) using a focused ion beam (FIB). The simple form of
the attack demonstrated here recovers the bitwise OR of two physically adjacent memory rows sharing common metal 1
contacts, however we have identified several potential mechanisms by which it may be possible to read the even and odd
rows separately.

We demonstrate the attack on a commodity microcontroller made on the 40nm node and show how it can be used to extract
significant quantities of sensitive data, such as keys for firmware encryption, in time scales which are very practical
for real world exploitation (1 day of sample prep plus a few hours of FIB time) with only a single target device
required after initial reconnaissance has been completed on blank devices.
\end{abstract}

\section{Introduction}

Microprocessors and low-cost microcontrollers often boot from external nonvolatile storage devices, such as QSPI NOR
flash, eMMC, or SD cards. This enables nonvolatile memory capacity to be easily scaled up or down to suit application
requirements, and reduces the cost of the processor by eliminating the silicon area spent on memory and extra
fabrication steps (eleven additional masks~\cite{1309900} in a 90nm process) required to fabricate Flash memory.

A major downside of this architecture from a security perspective, however, is that the contents of this memory are
easily accessible to an attacker wishing to reverse engineer the firmware or extract secrets (encryption keys,
certificates, etc.) from it. External memory devices can be desoldered and dumped using programming tools ~\cite{vnr,
xeltek}, and even in-package stacked die QSPI flash can be dumped with relatively low cost equipment (partial
decapsulation followed by landing microprobe needles on bond wires/pads \cite{257192}, or full decapsulation and
removal of the original bond wires followed by re-bonding the flash die to a new package). Minimal circuit reverse
engineering is required because the bond wires between the dice provide an obvious choke point for the data bus - all
that is required is to identify the clock and chip select (immediately obvious by the periodic waveforms) and determine
the ordering of the four data lines (straightforward to do by trial and error).

To mitigate these issues, processor vendors typically offer various forms of ``secure boot". Details vary from device
to device, however in the microcontroller space this commonly involves storing a signed and encrypted firmware image in
the external memory. A ROM bootloader verifies this image against a signing key stored in an on-die root of trust
(RoT), then decrypts it to on-die SRAM using a symmetric key stored in the RoT and executes it.

The RoT needs some way to store the signing key and decryption key. High volume applications sometimes use mask
ROM~\cite{scire2018attacking} as this is the lowest cost, but its inability to be modified after silicon fabrication
limits its usage to game consoles and other specialized environments in which the silicon manufacturer is the only
entity authorized to sign firmware. Additionally, mask ROM is well known to be readable by invasive attacks (deprocess
to the metal or via layer storing the data, image via optical or electron microscope as needed, and use machine vision
software to convert the images back to a memory dump). Mask ROM extraction, especially on older nodes, is being
performed at scale by vintage video game enthusiasts and archivists seeking to emulate classic games ~\cite{tms,
gameboy}, using software tooling widely available in the open source community, for example ~\cite{gerlinsky, rompar}.

In especially high-security applications, battery-backed SRAM is used since it can be zeroized by a self-destruct
feature in the event of tampering~\cite{ug570}.

For most applications, however, a field programmable nonvolatile memory is desired. One-time-programmable (OTP) fuse
memories are commonly used as modern fuse bitcell designs are compatible with unmodified CMOS processes
\cite{8388300, 1221214, son2011area, 1683903} requiring no additional mask steps to fabricate. An attacker wishing
to extract sensitive information from the external storage media attached to such a device must thus extract the
decryption key stored in the OTP memory.

The overall threat model for this work is an expert attacker with access to a SEM/FIB system and semiconductor
deprocessing equipment, who obtains physical access to their target and wishes to extract keys or other sensitive data
from an internal antifuse memory.

\section{Fuse Architectures}

Many different OTP fuse structures have been used over the years. Each has various tradeoffs in the design space in
terms of cost, area efficiency, compatibility with modern fabrication processes, and ease of data extraction by reverse
engineering techniques.

\subsection{Laser Fuses}

Laser fuses are simple metal links on the top metal layer which may be either left intact or cut by a laser cutter.
They are intended to be programmed at wafer test, before the devices are packaged. They do not scale well to modern
technology nodes~\cite[Page~40]{deloge2011analysis} due to the large size of the laser spot (limited by diffraction at
the laser wavelength) and the potential for molten metal splatter \cite[Figure~2.19]{deloge2011analysis}. Furthermore,
they cannot be field programmed since they require the laser beam to be able to reach the die surface, thus can only be
programmed before packaging.

No manufacturing process changes are needed at wafer fabrication time, since the fuses are simply strips of conductor
on the top layer. However, at programming time an expensive laser system must be available.

Due to their large size and easily accessible location on the topmost metal layer, laser fuse content can be trivially
extracted with only an optical microscope \cite[Figure~17]{skorobogatov2005semi}. No deprocessing is required.

\subsection{Metal Fuses}

Metal fuses work in exactly the same way as the fuses used in vehicles and homes for overload protection: a small metal
wire melts and ruptures when high current flows through it. These can be horizontal, using a simple metal
strip~\cite{1705255} as the fusible element, or vertical using a via~\cite{5437478}. These fuses, as with all of the
other non-laser designs, typically require a high voltage (higher than the normal device operating supply) to blow.
They may either use an internally generated charge pump or an external programming voltage, depending on the specifics
of the IP design and system requirements.

Metal fuses can be manufactured in an unmodified CMOS process, however some designs such as \cite{prom} use an extra
layer of high-resistivity metal such as nichrome. This allows higher density fuse structures at the cost of an
additional deposition, lithography, and etch step.

\begin{figure}
\begin{center}
\includegraphics[width=4cm]{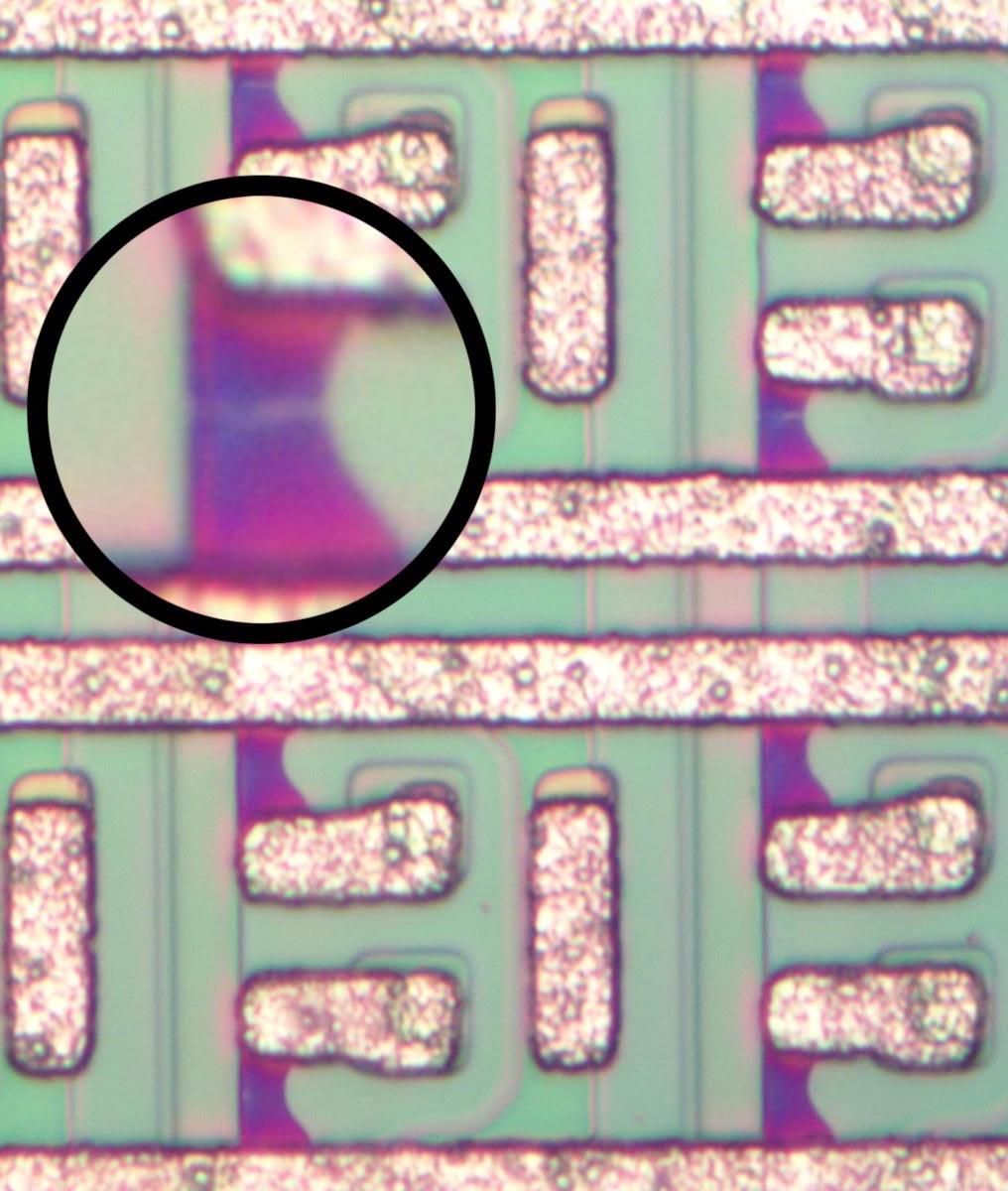}
\end{center}
\caption{\label{fig:metalfuse} Example of electrically programmed metal fuse on a 1970s-era PROM. The roughly 700nm
wide rupture in the nichrome fuse link is clearly visible under optical microscopy. Image from \cite{prom},
used with permission.}
\end{figure}

While the specific sample preparation required to image a metal fuse will depend on the bitcell structure, as a general
rule the ruptured region in a metal strip fuse is large enough - and the structural changes significant enough - that
bits are clearly distinguishable under scanning electron microscopy (SEM)~\cite{pufsec} once metal layers above the
fuse (if any) are removed by deprocessing techniques. On older nodes, blown bits may even be distinguishable under
optical microscopy (as seen in Fig. \ref{fig:metalfuse}) ~\cite{tl431, prom}.

\subsection{Silicide Electromigration Fuses}

Silicide electromigration fuses are similar to metal fuses at a high level, except that the fusible element is at the
polysilicon level~\cite{4405850, 1028987} rather than a metal layer. Passing a high current through the fuse link
causes electromigration, in which the silicide clumps up at the anode side of the fuse. The majority of current flows
in the silicide due to its lower resistivity, so there is minimal electromigration in the polysilicon and the ``blown"
fuse is not a complete open circuit, however the resistance increase from the ruptured silicide is sufficient to create
a clear distinction between blown and un-blown elements (e.g. $150 \Omega$ to $>3 k\Omega$ reported in \cite{4405850}).

Using polysilicon as the fusible element has significant benefits from a manufacturing perspective, since it is high
resistance (enabling small fuses) and already available in the standard CMOS process (no additional deposition or
lithography required).

These fuses are located on the polysilicon layer, so extracting data from them requires deprocessing. Once the poly is
exposed, blown fuses can be clearly distinguished under SEM by a region of missing silicide, as seen in
\cite[Figure~2]{4405850}, \cite[Figure~3]{4295075}. While it may be possible to read some bitcell designs on older
nodes optically, the fact that the poly is not ruptured - only the silicide - makes SEM a far better option. When using
a SEM to image in plan view, backscatter Z-contrast shows a clear difference between the silicide (containing a high
concentration of heavy metal atoms) and the polysilicon.

\subsection{Antifuses}

Antifuses, as the name implies, are the inverse of a fuse: the element starts in a high-resistance state and can be
irreversibly blown to a low-resistance state. Antifuses are typically based on a capacitor-like structure with two
conductors separated by a thin dielectric, providing isolation in the baseline (0) state. To program the antifuse, a
high voltage, on the order of 6 to 10.5V~\cite[Table~2.6]{deloge2011analysis}, is applied across the dielectric,
causing it to rupture and short the two conducting elements to each other.

\subsubsection{Metal-dielectric}

Early antifuses were located in the metal interconnect layers~\cite{BIRKNER1992561}, usually by adding a thin
dielectric between the top of a via and the bottom of the metal layer above it. This has fallen out of favor because it
requires an extra fabrication step to create and pattern the thin dielectric layer in the BEOL\footnote{Back End Of
Line - the dielectric and wiring layers above the active transistors}.

These antifuses, especially in older processes (for example, the 250nm Actel antifuses shown in ~\cite{antifusenotes})
have large enough feature sizes that dielectric breakdown is readily visible in SEM cross sections.

\subsubsection{Amorphous silicon}

Another antifuse technology, closely related to the metal-dielectric antifuse, uses a stack of titanium nitride (TiN)
and amorphous silicon (a-Si), between the top of a standard via plug and the bottom of the adjacent metal
layer~\cite{quicklogic}.

When programmed, the amorphous silicon recrystallizes and forms a low-resistance filament connecting the top metal
layer to the via. The a-Si layer and breakdown filaments are fairly large, meaning they can be easily imaged in SEM/FIB
cross section \cite[Figure~4]{quicklogic}.

As with the metal-dielectric antifuse, the amorphous silicon antifuse requires an additional deposition step to create
the a-Si and TiN stack, an additional lithography step to mask off the antifuse vias, and an additional etch step to
remove the a-Si everywhere except on the vias which will become fuses.

\subsubsection{Gate dielectric breakdown}

Modern antifuse structures typically are based on MOS transistors~\cite{1221214} and use gate dielectric as the fusible
element. This enables integration into an unmodified CMOS process with no extra deposition or lithography steps
whatsoever - a significant cost savings.

Due to the extremely thin dielectric (reaching 1 nm over 20 years ago~\cite{979590}) in the breakdown region, it is not
possible to image the broken-down gate dielectric in SEM (typical resolution limit 1-20 nm depending on settings and
instrument capabilities~\cite{resolution}).

The only technique capable of directly imaging gate dielectric breakdown is transmission electron microscopy (TEM).
Deliberately ruptured dielectric in antifuse bitcells would produce similar structural changes to those seen in
\cite[Figure~4.16]{tang2005physical}, however Tang's paper primarily focuses on dielectric breakdown as a device
failure mechanism rather than deliberate breakdown for fuse applications.

\subsubsection{Security of antifuses}

In an EDN technical article \cite{cunning}, DiPert admits that cross-sectioning individual bits is a plausible means of
extracting data from antifuses, and shows an example image of doing so by SEM cross section on an amorphous silicon
antifuse on a then-current process node\footnote{Layer thicknesses at the time were such that SEM had sufficient
resolution to see the breakdown. The switch to gate dielectric breakdown in modern fuse architectures, as well as the
continued scaling of layer thickness, means that SEM is no longer a viable option for imaging the breakdown region and
TEM must be used.}, but suggests that this is the only possible means of data extraction. The article makes somewhat
inflated estimates of the cost of performing an attack, assuming that an entire die must be destroyed to obtain a
single cross section of a fuse, leading them to conclude that dumping fuses from a die containing 800000 fuses will
require destroying 800000 dice. In reality, while cross sectioning fuses with a FIB may require destruction of a
\emph{few} adjacent bit cells, the true number of dice required for a full extraction by cross section is likely to be
a few dozen at most.

The cost of large scale data extraction by this mechanism, is, however, still likely to be prohibitive for most
attackers: performing a in situ cross section lift-out in a dual-beam SEM/FIB system \cite{langford2004situ} followed
by high resolution TEM imaging of each bit cell would take roughly a day of lab time, and several thousand dollars
(US) in equipment operating expenses and staff time \emph{per bit}\footnote{One commercial lab the authors contacted
quoted slightly over 3000 USD per specimen for in-situ FIB sample extraction and TEM imaging.} - or, at best, per string
of several adjacent bits within a single, roughly $10 \mu m$ wide, TEM lamella. Extraction of a 256-bit key by this
method, if the exact physical location of the key in the array is known, would thus cost in excess of half a million
dollars. If more samples are required in order to reverse engineer the memory structure, the cost would increase even
further.

Another effective (but very slow) technique is to use nanoprobes to directly measure resistance of each bit cell
\cite{son2011area, 1683903}. This will always be successful given sufficiently precise probing capability (since
there must be an electrically measurable change in the bitcell properties in order for the memory to function), but is
time consuming since bits must be measured one at a time. While far more affordable than cross sectioning for recovery
of small amounts of data such as keys at known locations, it is still impractical to use for dumping an entire fuse
array (making reverse engineering of the memory map challenging).

The difficulty and expense of directly imaging or probing the bit cells led IP vendors to make bold claims about the
security of antifuse memory, for example \cite{bitcell} claims ``State of memory (even for a few bits) is virtually
impossible to detect using physical attack or reverse engineering techniques". \cite{pufsec} claims "The burnout
creates a conductive path without leaving any visible traces on the surface" and shows metal/poly layer images of
programmed metal fuses and antifuses (the latter with no visible change between programmed and unprogrammed) under SEM.
Academic literature has also made similar claims of antifuse security, for example Skorobogatov states
\cite[page~28]{skorobogatov2005semi} ``it is virtually imposible to identify their state", giving ``an extremely high
security level".

The authors are also aware of at least one well known firm in the security testing space which had a general threat
model policy of assuming that contents of flash, ROM, or efuse could be easily extracted by an attacker, while contents
of antifuse memory were considered to be safe.

\section{Passive Voltage Contrast}

Voltage contrast in the FIB~\cite{campbell1995electrical} is a powerful technique for failure analysis which enables
visualization of logic states in an IC. The sample is scanned with the (positively charged) ion beam at low current,
which injects positive charges into the sample as well as releasing secondary electrons which may be used for imaging.

The secondary electrons are attracted back to or repelled from the sample surface depending on charges present on
various structures. Recaptured secondary electrons result in less secondary electron signal reaching the detector,
thus creating a voltage-dependent contrast mechanism in which more positive voltages appear dark and more negative
appear light.

Passive voltage contrast (PVC) involves using the imaging beam as the sole source of bias in the sample. Every pixel in
the region being imaged receives a uniform amount of charge per unit area from the scanning beam, but depending on
circuit connectivity some structures will dissipate the charge rapidly into the substrate while others will remain
charged for a longer period of time and show contrast as a result.

Active voltage contrast involves injecting voltages into the sample from an external source, either via the device pad
ring or by placing microprobes on circuit nodes. This is a more powerful technique but requires additional equipment
and setup time to utilize, thus PVC is often preferable if the necessary information can be gained.

While SEM PVC has been successfully used for direct readout of 350nm Flash memories\cite{courbon2016reverse} and FIB
PVC has been used for identifying pinhole defects in DRAM gate oxide \cite{rosenkranz}, the usage of PVC for readout of
antifuses has not been previously demonstrated in the literature.

\section{Test Subject}

Given that antifuses involve creating conductive paths to otherwise floating circuit elements, we hypothesized that PVC
would be an effective way to read out antifuse content in a rapid, scalable manner that did not require probing or
cross-sectioning every bit cell.

Lacking the funding and design expertise to create an antifuse test chip from scratch to test this hypothesis, we begin
searching for a commercially available device to use as a test subject for the research. Search criteria included:

\begin{itemize}
\item Readily available without NDAs or restrictions on use (ruling out many smartcard-type devices)
\item Contains antifuse memory which can be programmed by end users (ruling out devices which only use antifuses for
factory trim and calibration)
\item Antifuse array is relatively large, enabling multiple experiments per sample to reduce sample preparation time
\item Low cost (ruling out most FPGAs)
\end{itemize}

We immediately identified the newly-launched Raspberry Pi RP2350 microcontroller as a suitable candidate. It met all of
the requirements and was also the subject of a bug bounty competition~\cite{bounty1}, indicating that the device vendor
would likely be cooperative and willing to publish a security advisory if approached with a working attack.

\subsection{RP2350 Overview}

\begin{figure}
\begin{center}
\includegraphics[width=7cm]{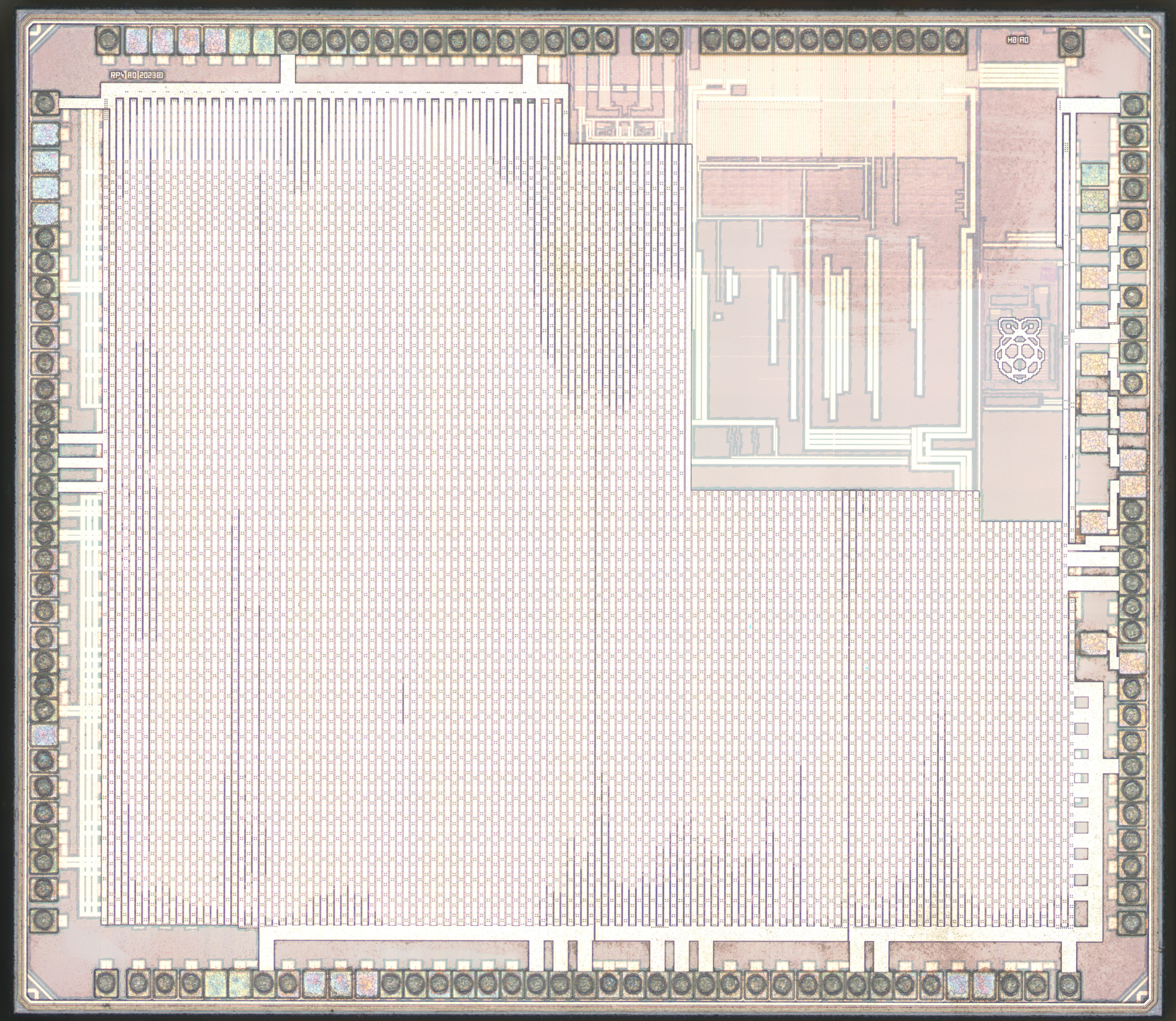}
\end{center}
\caption{\label{fig:top} Top metal of decapsulated RP2350 sample}
\end{figure}

\begin{figure}
\begin{center}
\includegraphics[width=7cm]{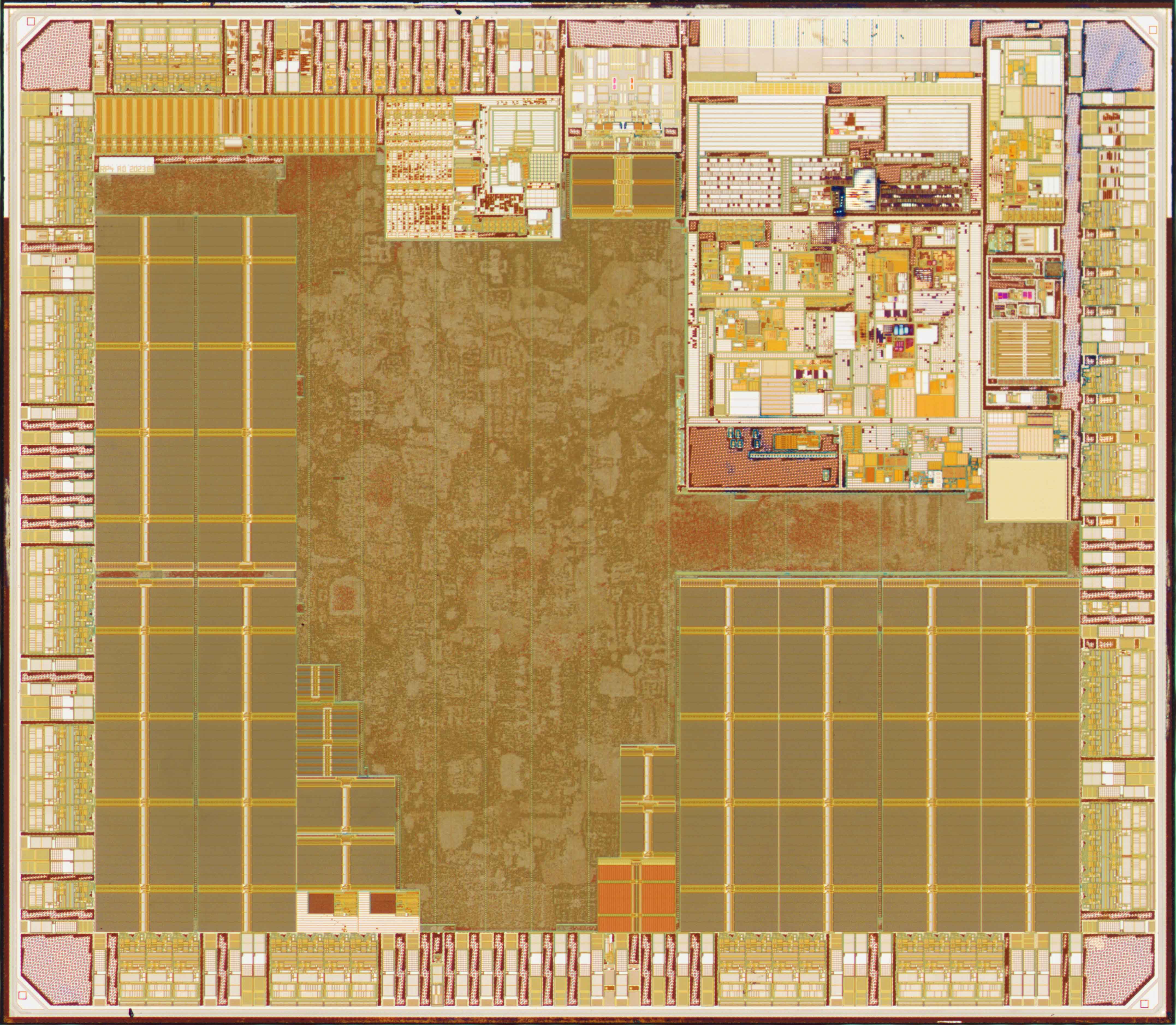}
\end{center}
\caption{\label{fig:substrate} Substrate view of decapsulated RP2350 sample with all metal and poly removed. The large
greenish memory arrays along the west side and in the southeast corner of the die are the main SRAM. Smaller green and
orange arrays in the south and north center are auxiliary SRAMs for peripherals, caches, and boot ROM. Fuses are the
yellow memory in the northwest corner.}
\end{figure}

The RP2350 (Fig. \ref{fig:top}, Fig. \ref{fig:substrate}) is Raspberry Pi's successor to their popular RP2040
microcontroller \cite{rp2350web}. It contains four 32-bit CPUs (two RISC-V and two ARM Cortex-M33, of which only two of
the four may be used at any given time), 520 kB of SRAM, various peripherals which are not relevant to this research,
as well as a 24 bit x 4K row antifuse memory array.

As is common for many modern low-cost microcontrollers, there is no on-die firmware flash. Firmware is loaded from
external QSPI flash~\footnote{Some RP235x SKUs offer stacked die in-package flash, but it is still a separate flash die
wirebonded to the same logic die as the no-flash version.} and either copied to the internal SRAM or executed in-place
(XIP) via the QSPI bus.

\subsection{Secure Boot}

Since the QSPI flash can be easily dumped or modified by an attacker with physical access to the board, the RP2350
supports a signed boot mode in which a binary stored in flash is copied to the internal SRAM, checked against a signing
key whose hash is burned into the fuses, and then executed if the signature check passes.

This can then be chained with a signed (but unencrypted) stub to create a full secure boot flow: the signed image
containing the stub and encrypted firmware is loaded into SRAM and integrity checked, then the stub decrypts the
encrypted blob using a key also stored in fuses.

The confidentiality of the firmware image, as well as any secrets stored in it, thus depends on the difficulty of an
adversary extracting data from the fuse memory.

\subsection{Synopsys 1T Bit Cell}

\begin{figure}
\begin{center}
\includegraphics[width=8cm]{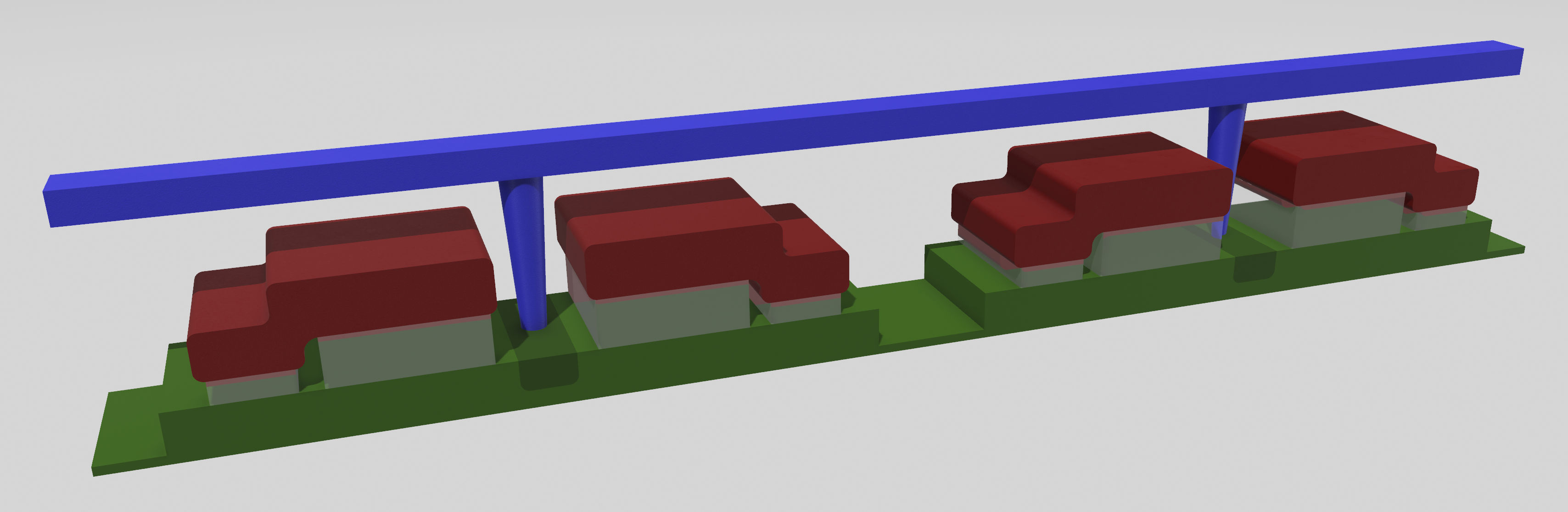}
\end{center}
\caption{\label{fig:bitcell} Simplified rendering of two rows of two bit cells (not to scale). Green = P substrate,
dark green = N+ implant, red = poly, blue = contact/metal 1. }
\end{figure}

The RP2350 fuse array uses the SHF NVM IP from Synopsys~\cite{datasheet}. The bit cell (Fig. \ref{fig:bitcell}) is
based on gate dielectric breakdown \cite{bitcell} and can be permanently blown from the default (logic 0) state to the
programmed (logic 1) state.

On the source side, the bit cell transistor contains a M1 contact (up to the bit line) and N+ implant, just like a
conventional NMOS transistor. The weak P channel is also largely the same as a conventional transistor in the I/O
cell, using thick ($V_{io}$) gate dielectric between the N+ doped polysilicon gate and channel.

The drain side however, lacks the expected N+ implant and M1 contact. Instead, the polysilicon gate overhangs the drain
end of the channel, but using thinner ($V_{core}$) gate dielectric. This effectively creates an NMOS transistor with a
dielectric between the drain (the overhanging region of poly) and channel, thus behaving as an open circuit regardless
of the control voltage on the gate.

\begin{figure}
\begin{center}
\includegraphics[width=8cm]{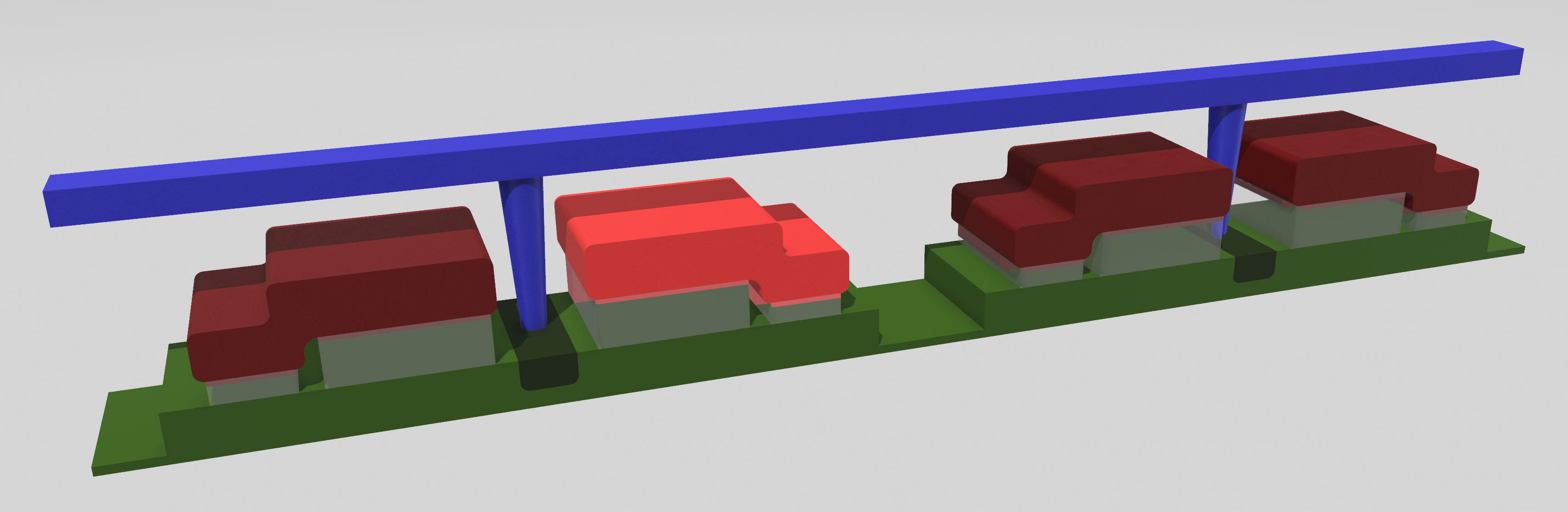}
\end{center}
\caption{\label{fig:bitcell3} Reading an unprogrammed bit cell. The polysilicon word line (red) is energized but the
fuse is not blown so no voltage is seen on the M1 bit line (blue). }
\end{figure}

\begin{figure}
\begin{center}
\includegraphics[width=8cm]{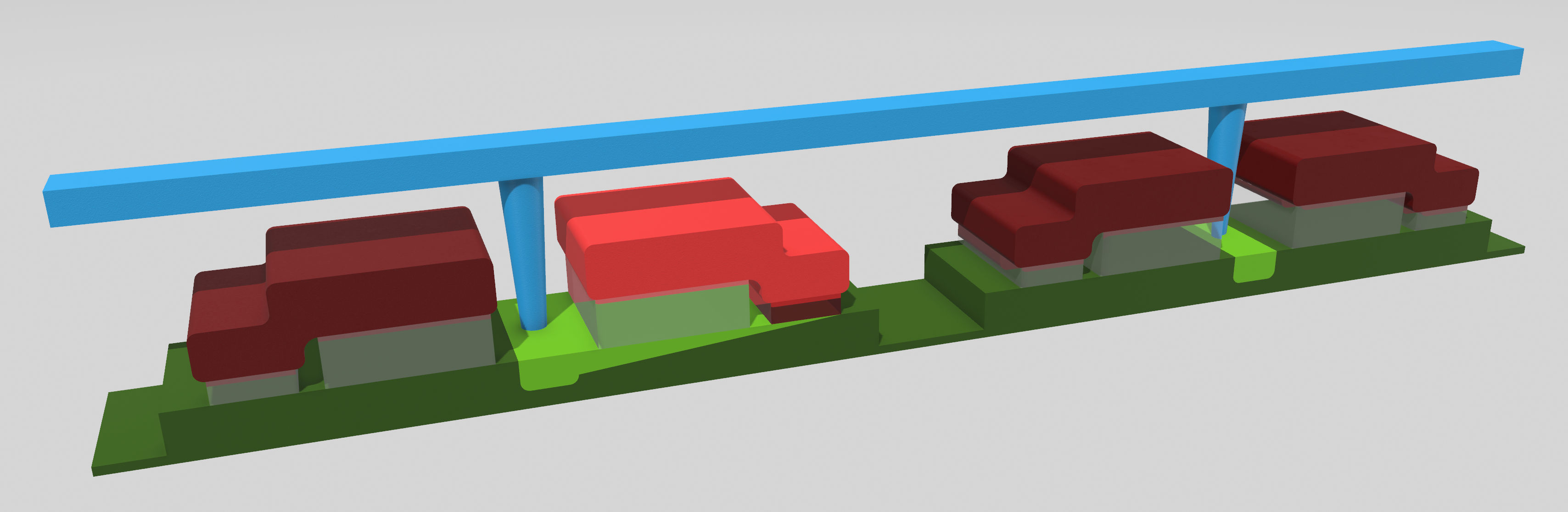}
\end{center}
\caption{\label{fig:bitcell4} Reading a programmed bit cell. The fuse element is blown so positive voltage on the word
line (red) conducts to the drain end of the channel (light green), through the channel to the source contact, and
energizes the bit line (light blue). }
\end{figure}

To read a fuse bit, the word line is driven high (turning on the select transistor), while the bit line is weakly
pulled low. If the fuse is not blown (logic 0) then the drain floats and the source and attached bit line remain low
(Fig. \ref{fig:bitcell3}). If the fuse is blown (logic 1) then the voltage on the gate travels from the drain through
the breakdown region into the channel, out the source, and brings the bitline high (Fig. \ref{fig:bitcell4}).

\begin{figure}
\begin{center}
\includegraphics[width=8cm]{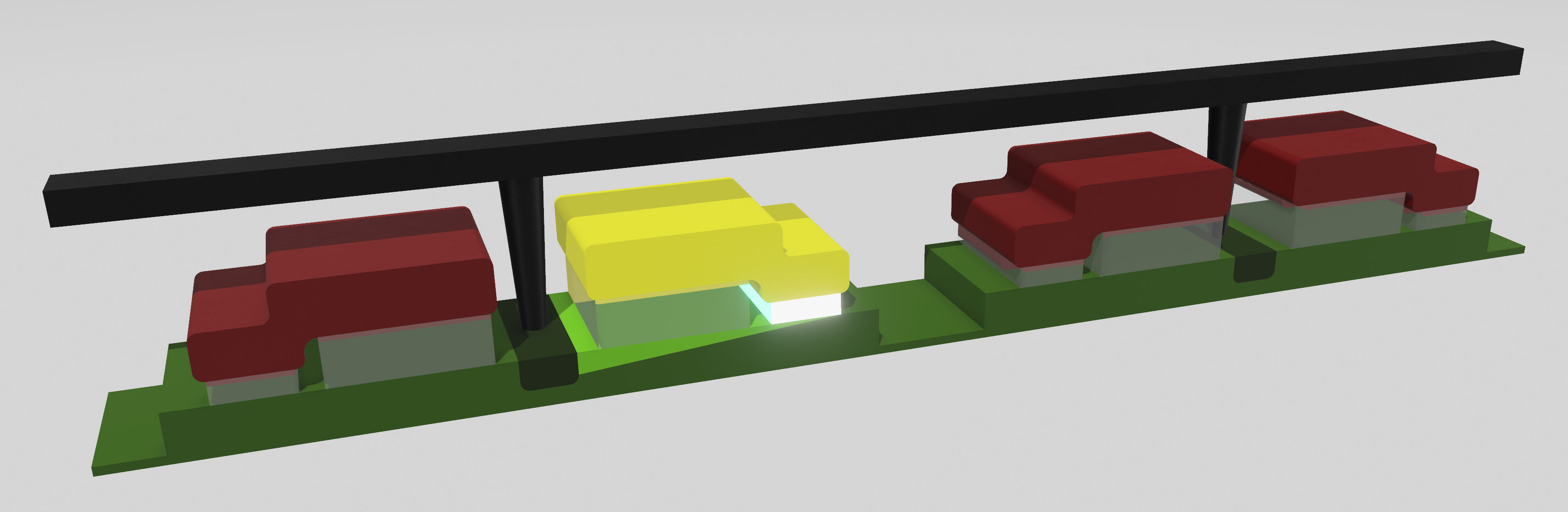}
\end{center}
\caption{\label{fig:bitcell5} Programming a bit cell. High voltage on the gate (yellow) and strongly grounding the bit
line (black) turns on the select transistor, applying a strong electric field across the thin dielectric and causing
it to rupture (white).}
\end{figure}

To program a bit (Fig. \ref{fig:bitcell5}), the word line is driven to a high voltage (supplied by an on-die charge
pump) while the bit line is driven strongly low. This results in the select transistor turning on, placing high voltage
between the channel (grounded) and drain (connected to the HV supply). The voltage is chosen to be high enough that the
thin $V_{core}$ dielectric between channel and drain will reliably rupture, while simultaneously low enough that the
thick $V_{io}$ dielectric between gate and channel is unharmed.

\subsection{Fuse Array}

\begin{figure}
\begin{center}
\includegraphics[width=8cm]{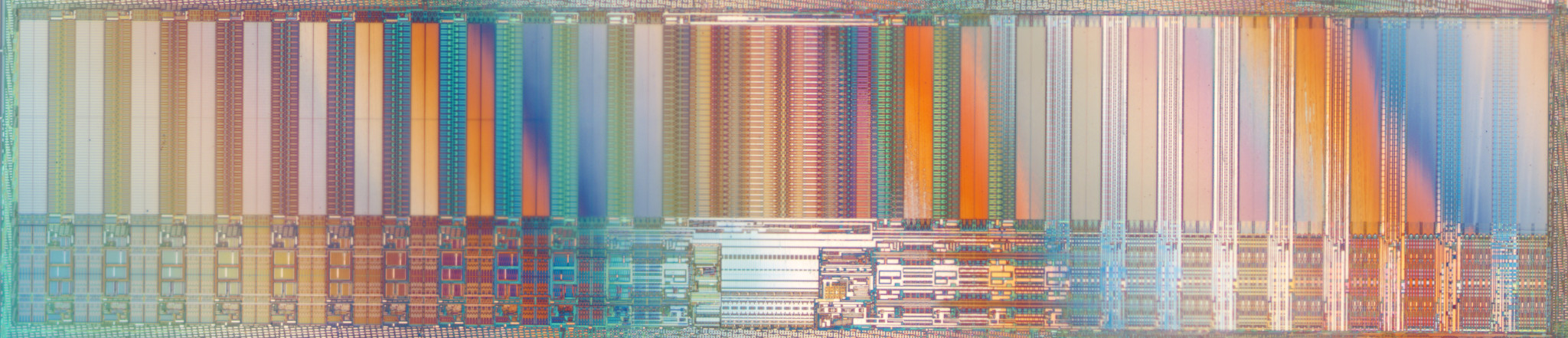}
\end{center}
\caption{\label{fig:fuses} Optical closeup of fuse array, bevel polish with M1-M4 visible left to right}
\end{figure}

The fuse array (Fig. \ref{fig:fuses}) is located in the northwest corner of the device, and consists of 24 bit planes,
each logically 1 bit x 4096 rows. The fuse memory can be accessed by the processor as either 4K 16-bit words with ECC,
or as 4K raw 24-bit words.

Supporting both modes allows each word of fuse data to optimize for data integrity (with ECC), or gain a 50\% increase
in memory capacity and allow multiple writes to the same fuse word (e.g. for device lifecycle management).

\begin{figure}
\begin{center}
\includegraphics[width=8cm]{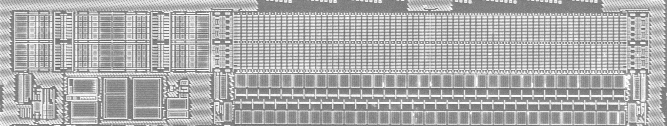}
\end{center}
\caption{\label{fig:single-col} SEM substrate view of single bit plane, rotated to better fit page layout.
North of die is to the right of the image.}
\end{figure}

Each fuse bit plane (Fig. \ref{fig:single-col}) consists of 2x2 identical bitcell arrays, each containing 18 columns of
34 rows of 2-bit unit cells. The outermost columns are dummy features for lithography purposes and are electrically
isolated, leaving 16 columns of active bit cells. The north and south row of each array contains hard-wired test
patterns used for internal consistency checks, leaving 16 x 32 (512) usable unit cells per array and 2048 unit cells
or 4096 bits per bit plane.

\begin{figure}
\begin{center}
\includegraphics[width=2cm]{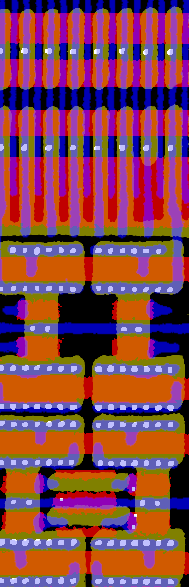} \includegraphics[width=2cm]{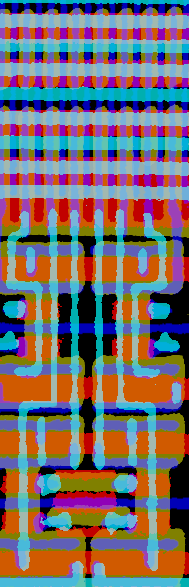}
\end{center}
\caption{\label{fig:layout} Vectorized layout of south end of a bit plane. Left: substrate (yellow), poly (red),
contact (white), M1 (blue). Right: contacts hidden, M2 (cyan) shown. }
\end{figure}

Poly word lines run east-to-west across the array (Fig. \ref{fig:layout}), while M1 bit lines run north-south over bit
cells. Additional metal strips run north-south between bit lines on M1, perhaps grounds for shielding although we did
not fully reverse engineer this part of the circuit.

The north and south bit cell arrays within a plane are identical tiles mirrored about the horizontal centerline of the
array. Addressing for the south tile increments from row 0 at the south end to row 31 at the center of the array, while
the north tile increments from row 32 at the north end to row 63 at the center.

\begin{figure}
\begin{center}
\includegraphics[width=6cm]{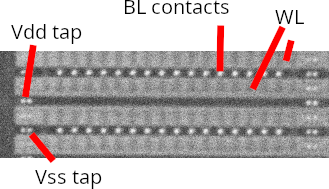}
\end{center}
\caption{\label{fig:poly} Bitcell array seen at poly/contact layer.}
\end{figure}

\begin{figure}
\begin{center}
\includegraphics[width=8cm]{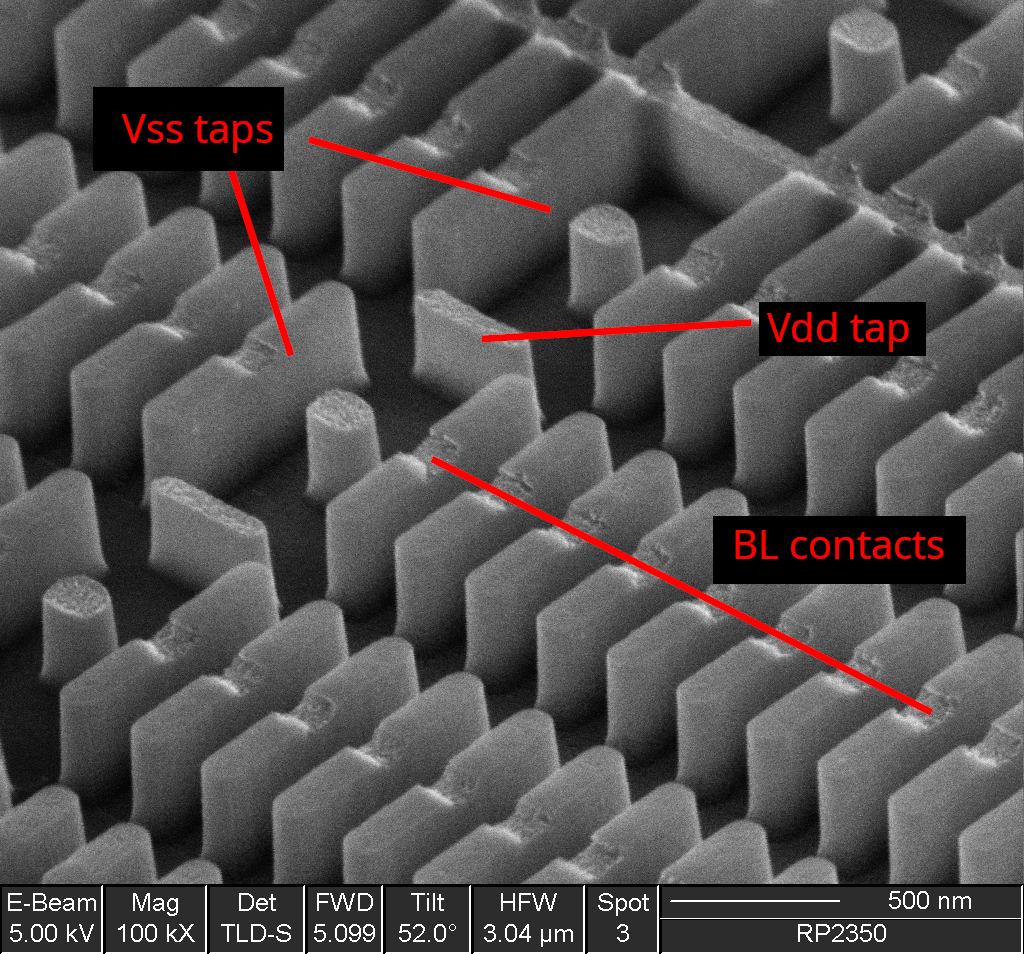}
\end{center}
\caption{\label{fig:well-taps} Angled SEM substrate view (STI oxide and poly removed) of well tap positions in relation
to bit cells}
\end{figure}

Looking at the poly/contact layer in Fig. \ref{fig:poly}, wordline contacts at each side of the bitcell string can be
seen going up to the parallel word line on M2. In between the wordlines, a row of sixteen bitcell contacts go up to the
M1 bitlines. Additional contacts for well/substrate taps can be seen going down to the substrate between the wordline
contacts. An angled substrate perspective (Fig. \ref{fig:well-taps}) shows the structure of the bitcells in relation to
the well taps clearly.

\section{Data Extraction}

\subsection{Rough Sample Preparation}

Each sample was decapsulated with fuming nitric acid according to industry standard processes, then cleaned with
acetone. The bare die was then mounted to a silicon carrier substrate with hot-melt adhesive for easier handling and
attached to a polishing fixture (15-1010, Allied High Tech, Cerritos, CA).

The fixture was leveled to achieve a parallel polish, and the die was mechanically polished~\footnote{During testing,
it was found that this fuse IP is sensitive to some other deprocessing techniques. For example, one test using 2 keV
argon ion milling gave an excellent surface finish but resulted in every antifuse in the array blowing to the ``1"
state. Based on this early result, plasma and ion beam deprocessing techniques were avoided for the remainder of the
research since we already had achieved good results with mechanical polishing. Further testing of process parameters is
needed to more accurately quantify the damage thresholds and determine if plasma deprocessing with different parameters
can be a viable technique for this IP or if exclusive use of wet chemical / mechanical techniques is required for bulk
material removal. FIB milling of small areas (e.g. single bit plane) does not appear to damage data.} using diamond
lapping films (50-30075, Allied High Tech, Cerritos, CA) until the fuse array had metal 1 fully exposed and reduced in
thickness, but still visible optically. The sample was then cleaned with water to remove polishing debris and
lubricants and blown dry with compressed air.

\subsection{Final Preparation and Imaging}

After cleaning and drying, samples were loaded in a dual beam SEM/FIB (Strata 235, FEI Company, Hillsboro, OR) and
prepared for final milling.

One bit plane at a time was milled using the Ga+ beam only, with no gas chemistry, at 300 pA to slowly remove the
residual metal 1 and ILD and expose the contacts. SEM images were taken periodically during the milling operation in
order to avoid over-milling once M1 was removed.

After contacts were clearly visible and all M1 was gone, a short cleanup milling pass at much lower (10 pA) beam
current with $XeF_2$ enhanced etch chemistry was performed to remove any sputtered debris which might have redeposited
onto the bit plane during the rough FIB mill.

\begin{figure}
\begin{center}
\includegraphics[width=8.5cm]{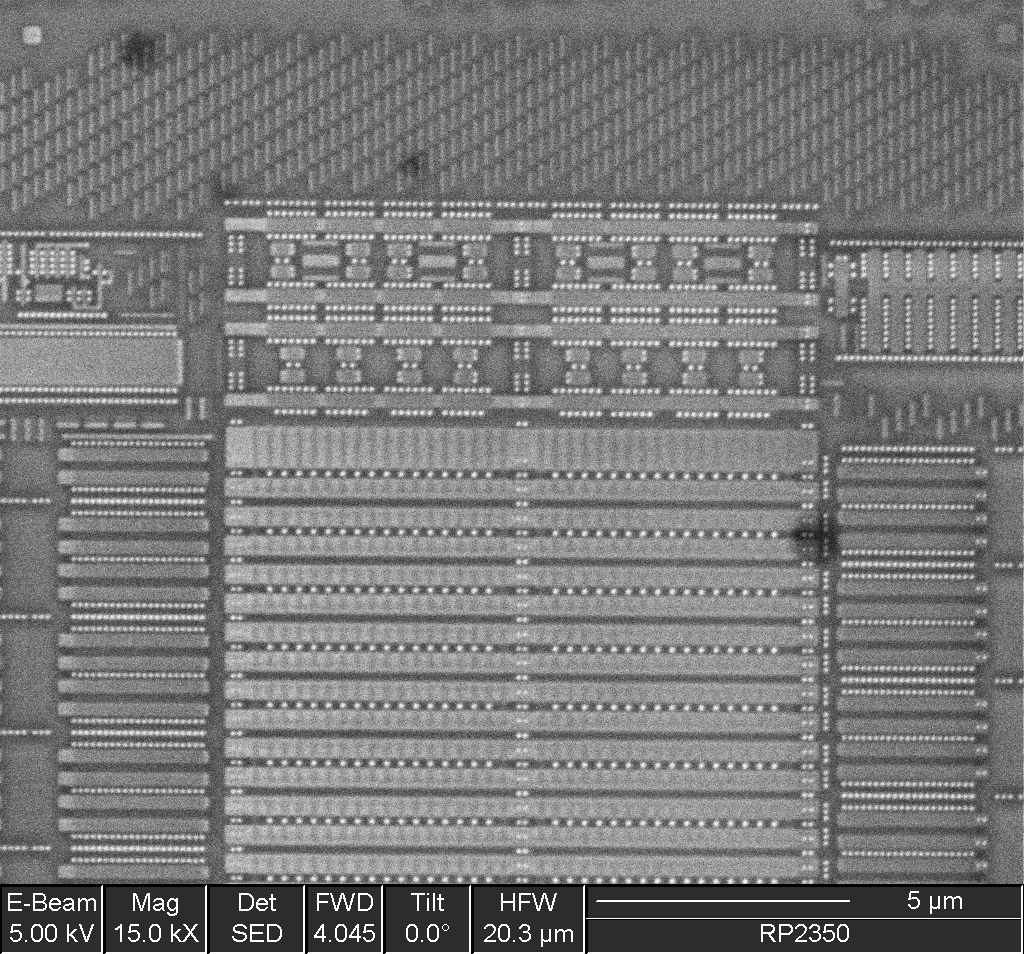}
\end{center}
\caption{\label{fig:sem-contact} North end of fuse bit plane after deprocessing, as seen in SEM.
No voltage contrast is visible since negative bias induced by electron beam turns all select transistors off. }
\end{figure}

At this point the sample is ready for analysis, but SEM imaging (Fig. \ref{fig:sem-contact}) shows no voltage contrast.

This is because the wordlines are all negatively biased by the electron beam so the NMOS bitcell transistors are
switched off (gate at or below substrate potential) and thus no current can flow through the fuses. All bitline
contacts have roughly the same leakage resistance into the substrate, so they all charge to the same negative voltage
under the electron beam irradiation and no PVC occurs.

Using the positively charged ion beam, the wordlines will charge to a positive voltage and the bitcells will turn on,
causing contrast between programmed and unprogrammed cells to appear.

\begin{figure}
\begin{center}
\includegraphics[width=8.5cm]{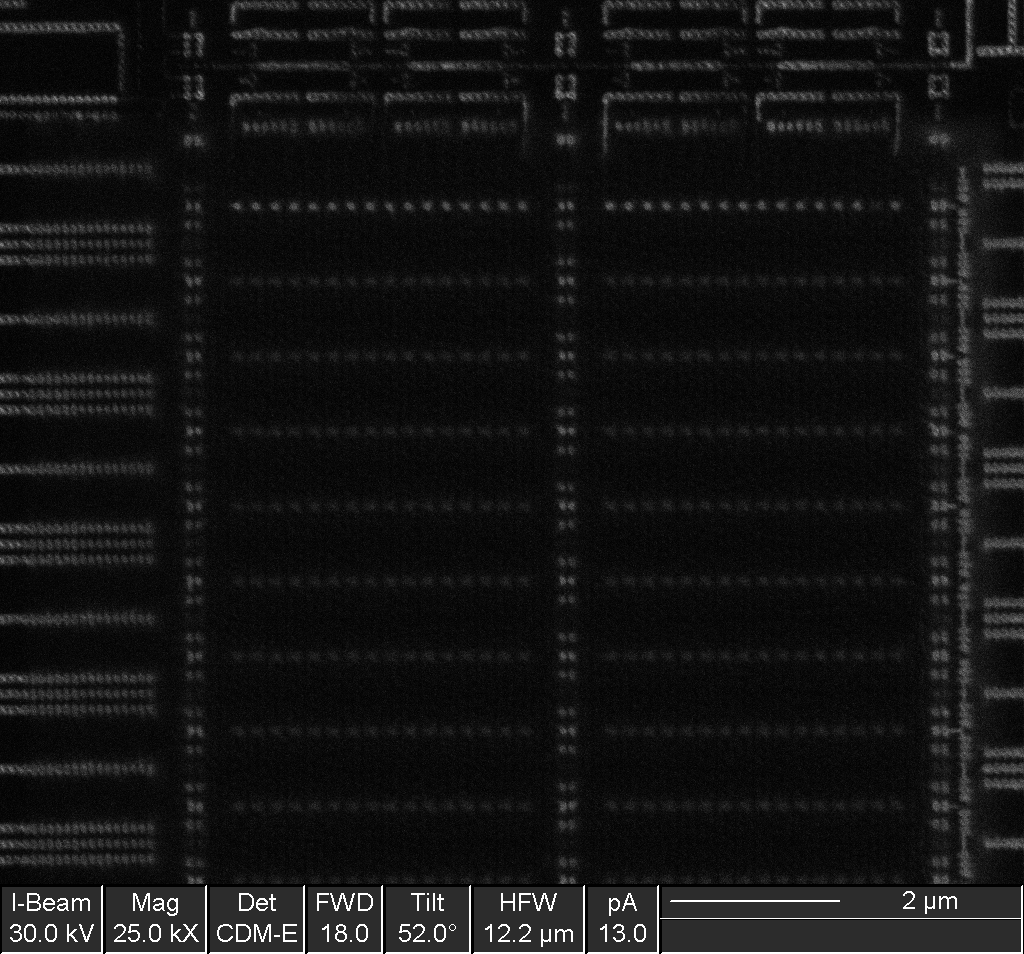}
\end{center}
\caption{\label{fig:charged} North end of fuse bitplane showing poor PVC image due to positively charged sample.
Image is dim despite detector contrast set to maximum.}
\end{figure}

Milling operations with the ion beam, especially at higher beam current, tended to result in the sample surface
becoming positively charged. This resulted in a general reduction in contrast and darker image (Fig.
\ref{fig:charged}), as negatively charged secondary electrons from the sample were attracted to and recaptured by the
positively charged sample.

\begin{figure}
\begin{center}
\includegraphics[width=8.5cm]{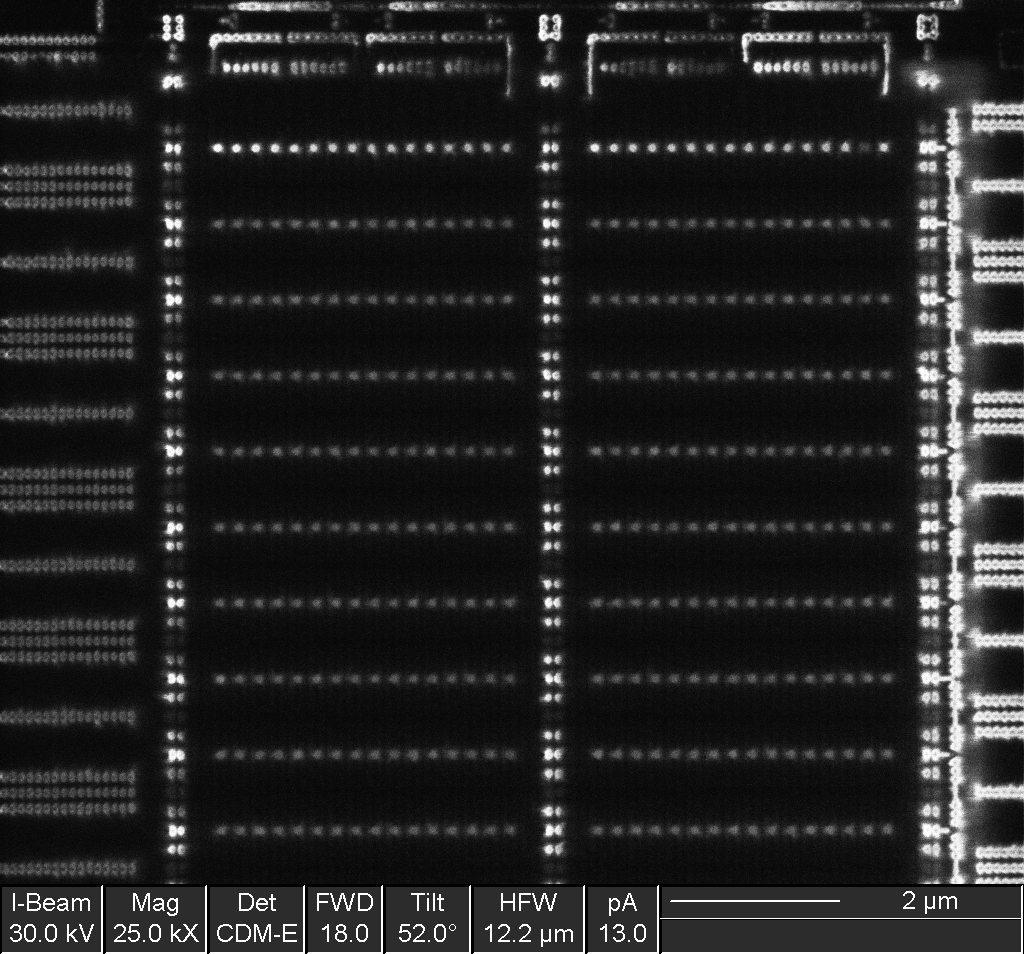}
\end{center}
\caption{\label{fig:discharged} Same region of sample after using electron beam to neutralize positive charge buildup.
Detector contrast is reduced below maximum to improve image quality due to the significant increase in secondary
electron signal.}
\end{figure}

Some FIBs are equipped with a low energy flood electron gun~\cite{barnard2006flood} to provide charge neutralization
however the authors' system lacks this capability. We instead neutralized charge on the sample manually with the
electron column, by periodically performing a slow scan of the region of interest. This resulted in a marked increase
in PVC image brightness and overall quality (Fig. \ref{fig:discharged}).

\subsection{Data Analysis}

Images were manually registered due to the small number of FIB images (approximately ten tiles on our system, depending
on overlap percentage) required to cover each bit plane.

\begin{figure}
\begin{center}
\includegraphics[width=8cm]{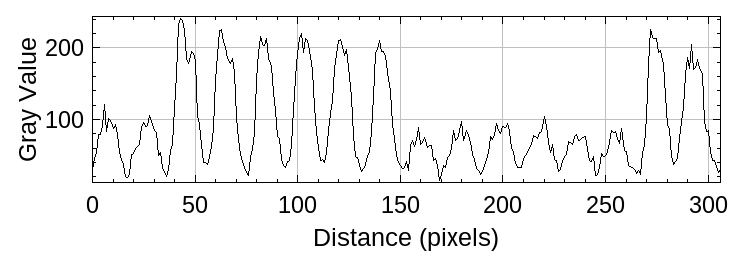}
\end{center}
\caption{\label{fig:contrast} Plot of PVC signal strength (arbitrary units) vs distance across one column of bits,
showing significant contrast between programmed (strong signal) and unprogrammed cells (weak signal).}
\end{figure}

Analysis of the captured images with FIJI~\cite{schindelin2012fiji} shows strong contrast - over 100 gray levels -
between dark (logic 0) and light (logic 1) vias (Fig. \ref{fig:contrast}), enabling highly reliable automatic data
extraction by machine vision tools.

Our initial hypothesis for PVC was that the positively charged ion beam would result in the wordlines being biased
high, turning on the bit cells and allowing positive voltage from the wordline to reach the bitline contacts if either
cell sharing the contact was programmed. Unprogrammed bits would cause the via to remain undriven, bleeding off
injected charge to the substrate. This would, however, cause programmed bits (high voltage) to show up darker than
unprogrammed (low voltage); the data showed the opposite.

\begin{figure}
\begin{center}
\includegraphics[width=8.5cm]{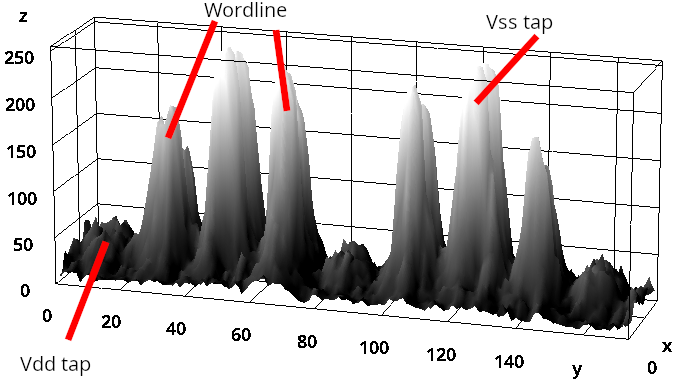}
\end{center}
\caption{\label{fig:pvc-voltages} Surface plot of PVC signal intensity (arbitrary units) vs X/Y position (pixels) on
wordline/well contacts. Lighter colors and higher peaks indicate stronger PVC signal (more negative voltage). }
\end{figure}

Closer analysis of brightness values across the wordline and well contacts (Fig. \ref{fig:pvc-voltages}) provided an
answer: the wordline voltage is actually fairly low, but noticeably higher than the Vss taps.

\begin{figure}
\begin{center}
\includegraphics[width=8.5cm]{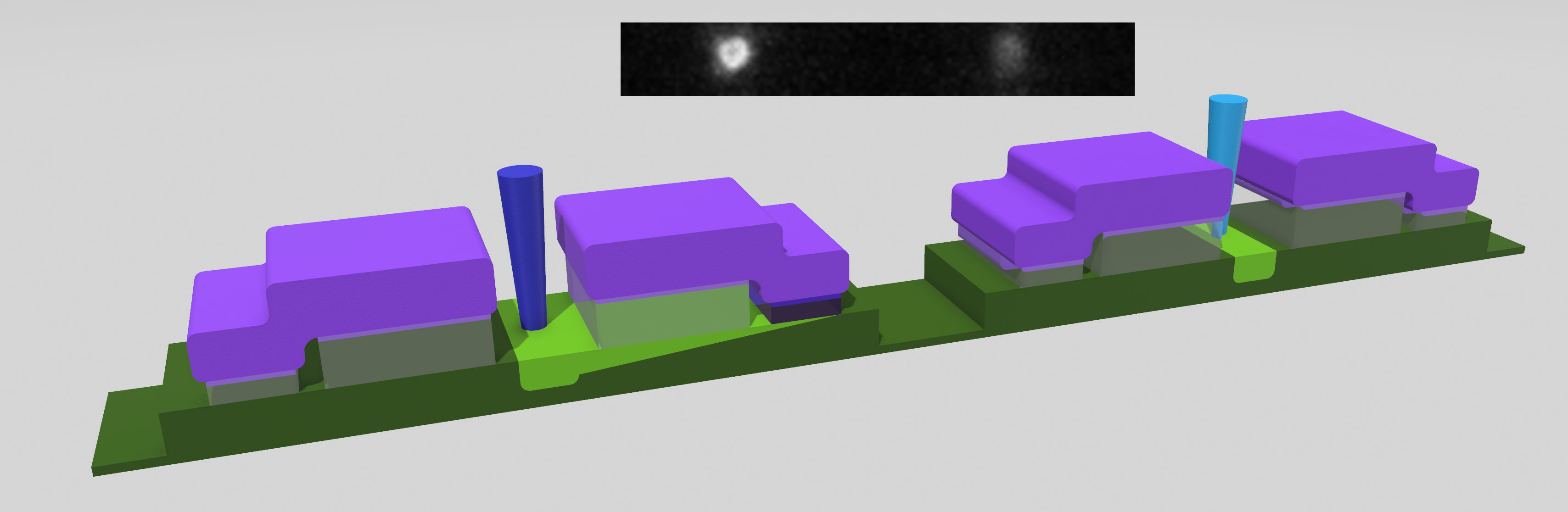}
\end{center}
\caption{\label{fig:pvc-overlay} Bit cells during PVC readout. Wordlines (purple) are biased to approximately $V_t$ by
equilibrium between 10 pA injected beam current and wordline leakage to ground. Right bitcell in the left pair has the
fuse blown (black), so current injected into the left bitline contact can dissipate to ground through the transistor
channel (light green), keeping the bitline contact at a low voltage (dark blue) and producing a strong PVC signal as
seen in the FIB image inset. No bitcells in the right pair are blown, so injected current has no escape path and the
bitline contact charges to a high voltage (light blue), producing a weak PVC signal.}
\end{figure}

We propose that this is caused by beam-induced charge biasing the wordline up to approximately $V_t$ of the bitcell
transistors, but being prevented (by antenna diodes and/or wordline driver transistor leakage paths) from going any
higher. Since the bitcell transistors are conducting weakly, charge deposited on the bitline vias of programmed cells
(logic 1) will be dissipated to the substrate, causing the vias to show up as low voltage (strong PVC signal). If the
cell is unprogrammed (logic 0), however, the via will build up a strong positive charge without any easy leakage path
to ground, causing it to show up as high voltage (weak PVC signal), as seen in Fig. \ref{fig:pvc-overlay}.

\subsection{Address Map Recovery}

Test devices were programmed with unique patterns in order to aid in reverse engineering the address map. The first
test sample used a simple repeating pattern of 1-0-1-0, 1-1-0-0, 1-1-1-1-0-0-0-0 bits. Subsequent samples included test
patterns such as per-bitplane unique values, horizontally and vertically asymmetric patterns to detect mirroring, etc.

Reversing the full address map took only two test devices and approximately three days of lab time including sample
preparation. We then created a Python script which could convert from a hex dump of fuse values to an ASCII art render
of the entire fuse memory and validated that the rendered fuse maps matched observed PVC extractions.

\begin{figure}
\begin{center}
\includegraphics[width=3.8cm]{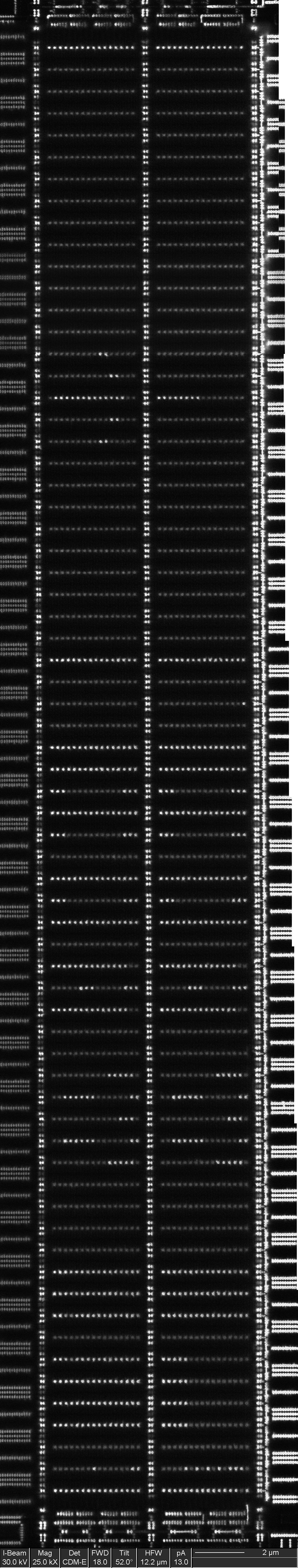}
\end{center}
\caption{\label{fig:full-bitplane} Test extraction of an entire antifuse bitplane (2048 bitcell pairs). }
\end{figure}

To demonstrate our attack scales to larger quantities of data we extracted the content of an entire bitplane (4096 bits
as ORed pairs, giving 2048 bits of leaked data), seen in Fig. \ref{fig:full-bitplane}, in approximately one hour,
including the final FIB milling pass but not decapsulation or the mechanical polish to metal 1. The test pattern
includes various asymmetric features used for address map reversing, a pixel-art "happy cat face" =3\footnote{A popular
text emoji in internet culture \url{https://knowyourmeme.com/memes/3-cat-face}} doodle, an arrow pointing at the
address of the bug-bounty key, and initials of our institution (redacted for review). The lowest two rows of data are
the factory trim / serial number data (fuse page 0) and the internal calibration row (lowest row, not accessible via
the normal address map) and are not part of the test pattern we programmed.

\section{Future Research}

Our current attack only recovers the bitwise OR of the two bitcells sharing each contact and cannot distinguish between
the north fuse, south fuse, or both being blown.

We conjecture that it may be possible to optimize the PVC technique in order to determine the Hamming weight of the
paired bitcells (i.e. distinguish the both-blown case from one blown) by adjusting beam parameters and detector
contrast so that the both-blown case produces a stronger PVC signal (higher leakage current) than one-blown however
attempts to date have not been successful.

We also intend to further explore various techniques for extracting north or south halves independently. Methods under
consideration include reduced-area scans (so that only one wordline is biased positively and the adjacent bitcell
transistor remains off) and using FIB metal deposition to short one wordline to ground while the other is left to float
high under ion beam irradiation.

\section*{Conclusions}

We have demonstrated that FIB passive voltage contrast is a straightforward and effective technique for extraction of
significant amounts of key material or other sensitive data from the Synopsys SHF NVM on the TSMC 40 nm process. The
device physics exploited by the attack have broad applicability to antifuse memories across a wide range of foundries,
technology nodes, and bitcell designs. Further research is warranted to refine the technique and explore which types of
memories can and cannot be extracted by this method.

The attack demonstrated here only recovers the bitwise OR of the low and high halves of a 64-word memory page, however
given the RP2350's page locking scheme (in which permissions are applied separately to each page) it is very likely
that many pages in real-world deployments will only store a single key with the remainder of the page left unprogrammed
(logic 0), leading to complete recovery of the key. Additionally, array structures in which a bitcell contact is
dedicated to a single bitcell can be fully extracted by our technique.

While a FIB system is a very expensive scientific instrument (costing several hundred thousand USD, plus ongoing
operating expenses in the tens of thousands per year), it is possible to rent time on one at a university lab for
around \$200/hour for machine time\footnote{For example, the University of Washington currently charges commercial users
\$200/hr for FIB time plus \$165/hr for staff assistance} or around twice this for machine time plus a trained operator
to run it. This is low enough to be well within the realm of feasibility in many scenarios given the potential value of
the keys in the device.

Designers and users of semiconductor devices whose security model is based on the premise that antifuse memory is
extremely difficult for an adversary to dump should perform a risk assessment to determine their exposure.

We have identified a near term mitigation for our attack against this specific bitcell array structure. Programming the
opposite side of each bitcell pair to the complement of the data bit - in other words, bit position N should be set in
exactly one of fuse word M or word (M xor 32) - will ensure that the basic PVC technique demonstrated here reads the
data as all 1s. Given the multiple techniques we are exploring to read even/odd wordline data separately, it is likely
possible to defeat this mitigation, but it is a low-effort means to increase the difficulty of an attack and does not
require a silicon spin.

As a general rule, the authors do not recommend assuming that \emph{any} semiconductor memory cannot be dumped by an
adversary. Memories can be easily recognized by a reverse engineer due to their regular structure and easily stand out
as targets which may contain interesting data.

Encrypting or obfuscating sensitive data stored in nonvolatile memory, using a cryptographic engine built out of
autorouted standard cell logic, significantly raises the difficulty of an attack since the adversary must identify and
reverse engineer the crypto circuit in a sea of millions of gates without any obvious clues to point to its location.

An even more effective protection is to use per-device keying and/or asymmetric cryptography to completely eliminate
the need for storage of high value, long-lived secrets in nonvolatile memory.

\section*{Ethical Considerations}

This research was performed as a submission to a bug-bounty competition organized by the silicon vendor, Raspberry Pi,
in which the security community was openly invited to attack the RP2350 and find ways to extract secrets from the
fuses.

Full details of the vulnerability were disclosed to the silicon vendor in September 2024 as part of our submission to
the competition (and to Synopsys, the IP vendor, by Raspberry Pi). A description of the attack was released to the
public in January 2025 as part of a vendor security advisory~\cite{bounty2}. We coordinated our submission of this
paper with the vendor to minimize impact to them and their customers.

While this attack enables extraction of secrets, such as firmware decryption keys, stored in RP2350 devices, we believe
that raising awareness of this weakness will help to improve security across the semiconductor industry and that this
benefit outweights the downsides. Notably, this is \emph{not} a secure boot bypass and does not enable mass compromise
of already-shipped hardware protected by the RP2350's secure boot. Additionally, a fault injection attack (discovered by
another individual participating in the same competition) enabling extraction of fuse content from the RP2350 was made
public in December~\cite{cullen}. Our work had already been completed by this point, but the existence of another
public attack vector further tipped the balance in favor of full disclosure of our technique.

From a broader industry perspective, many other devices besides the RP2350 are likely susceptible to similar attacks
and we believe enabling other teams to reproduce this work will enable appropriate testing to be performed so
mitigations can be put in place as required.

We chose not to anonymize the target device or fuse IP bcause the security advisory published by Raspberry Pi lists the
RP2350 as an affected device, and the RP2350 datasheet openly states that the fuse IP is the Synopsys SHF NVM. Thus,
redacting identifying details of the device or IP in this paper would serve no purpose. Clearly stating the tested
product will enable users to more easily determine their risk.

Reverse engineering the physical address map of the memory took the authors only three days. This is not a significant
impediment to a real world attacker, so we chose to release the address map in the spirit of the open science policy to
ease replication of our work.

\section*{Open Science}

This paper describes the PVC fuse extraction technique and sample preparation in sufficient detail to enable other
research groups to replicate the work. Key microscope configuration parameters are included in the image databars.

The full physical address map of the RP2350 is included in the appendix, enabling other groups to easily program test
devices with test patterns of their choice and experiment with data extraction techniques.

A series of Python scripts for converting a linear fuse dump from the "picotool" utility to a physically addressed
ASCII art render (of both the individual bit values and the OR'd values seen via PVC), as well as for converting a
desired test pattern to a linear fuse map, have been uploaded to an anonymous pastebin for review. The camera-ready
version of the paper will link to a more permanent GitHub repository or similar.

\begin{itemize}
\item burn\_fuses\_from\_file.py (writes fuse database to a chip): \url{https://pastebin.com/0VQZ9eBA}
\item fuse\_from\_render.py (converts ascii art to memory dump): \url{https://pastebin.com/bKa8Ei7W}
\item render.py (converts memory dump to ascii art): \url{https://pastebin.com/haMzHsVp}
\end{itemize}

\section*{Acknowledgments}

The authors would like to thank Raspberry Pi for their cooperation throughout the competition and disclosure process,
as well as Entropic Engineering for assistance with procuring scarce RP2350 samples shortly after the device had been
released.

\bibliographystyle{plain}
\bibliography{usenix2025-antifuse}

\appendix
\section{RP2350 fuse physical address map}

\begin{figure}
\begin{center}
\includegraphics[width=9cm]{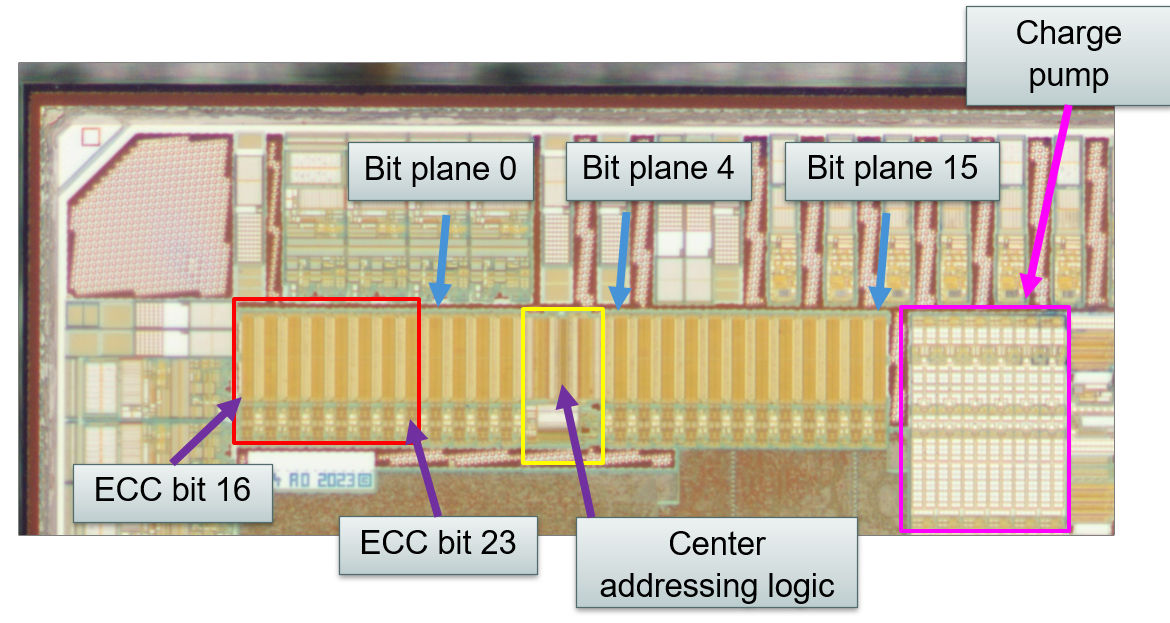}
\end{center}
\caption{\label{fig:bitplanes} Bit plane ordering for RP2350 fuse memory}
\end{figure}

The RP2350 fuse IP is organized as 24 bit planes, 12 on either side of a central address spine (Fig.
\ref{fig:bitplanes}). Starting from the west, the bit ordering is ECC bits (16 to 23), data bits 0 to 3, address logic,
data bits 4 to 15.

\begin{figure}
\begin{center}
\includegraphics[width=9cm]{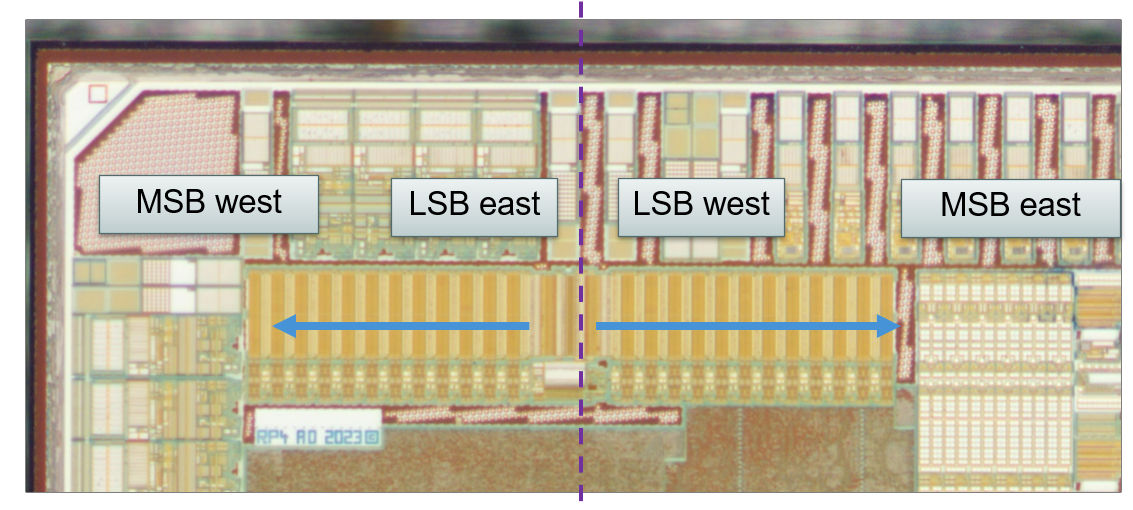}
\end{center}
\caption{\label{fig:mirror} Column mirroring and overall address ordering}
\end{figure}

The physical layout and column addressing for the east half (columns 4-15) is mirrored relative to the west half (Fig.
\ref{fig:mirror}); all column addressing and layout discussed in this appendix is oriented for a column in the west
half of the array.

\begin{figure}
\begin{center}
\includegraphics[width=9cm]{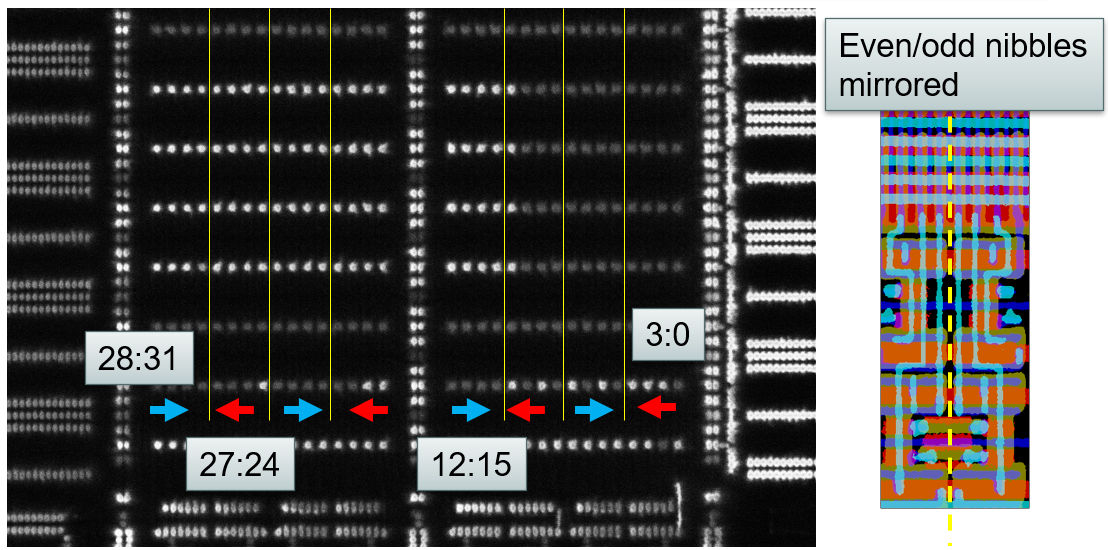}
\end{center}
\caption{\label{fig:colmap} Column address ordering for west half of the array (ECC bits and data bits 0-3). East half
of the array mirrors the entire column left-to-right. Arrows denote order of increasing address within the 4-bit group.}
\end{figure}

Within each bit plane, data is organized in two columns (Fig. \ref{fig:colmap}), each consisting of four groups of four
bits. Column addresses increment broadly from right to left, however odd-numbered groups have bit addressing from left
to right within the group while even numbered increment from right to left.

\begin{figure}
\begin{center}
\includegraphics[width=9cm]{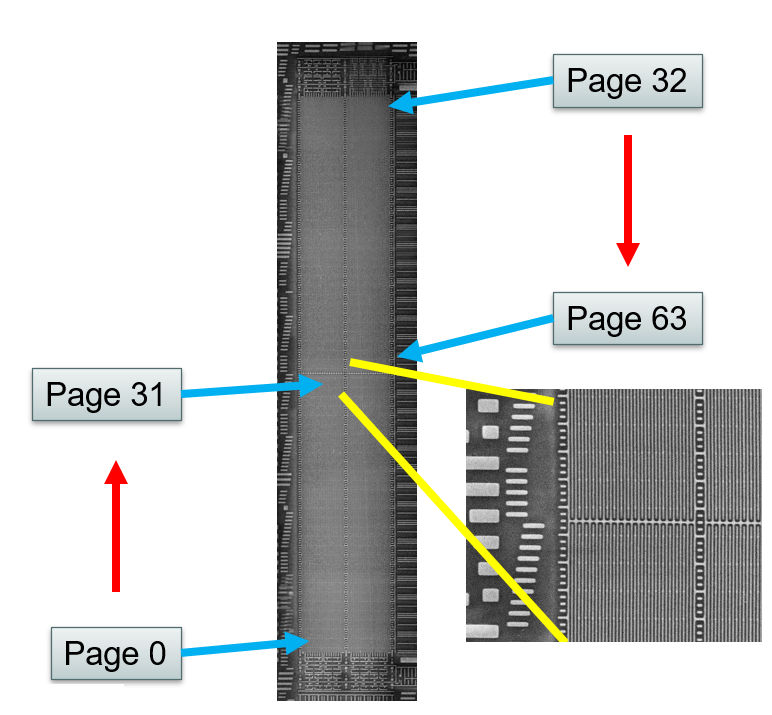}
\end{center}
\caption{\label{fig:rowaddr} Row addressing for a single RP2350 fuse bit plane}
\end{figure}

Row addressing starts from the south end of the array (Fig. \ref{fig:rowaddr}) with fuse page 0 located just north of
the calibration row and page 31 just south of the centerline. The north half of the array is vertically mirrored
relative to the south, with page 32 at the far north end and page 63 just north of the centerline.

\end{document}